%% file: main.tex
\documentclass{article}

\usepackage{arxiv}

\usepackage[utf8]{inputenc} 
\usepackage[T1]{fontenc}    
\usepackage{hyperref}       
\usepackage{url}            
\usepackage{booktabs}       
\usepackage{amsfonts}       
\usepackage{nicefrac}       
\usepackage{microtype}      
\usepackage{cleveref}       
\usepackage{lipsum}         
\usepackage{graphicx}
\usepackage{natbib}
\usepackage{doi}
\usepackage{multirow}
\usepackage{threeparttable}

\title{Predicting Axillary Lymph Node Metastasis in Early Breast Cancer Using Deep Learning on Primary Tumor Biopsy Slides}

\author{
	Feng Xu$^{1\ast\dag}$
	\hspace{0.15in} Chuang Zhu$^{2\ast\dag}$
	\hspace{0.15in} Wenqi Tang$^{2\ast}$ 
	\hspace{0.15in} Ying Wang$^3$ 
	\hspace{0.15in} Yu Zhang$^2$ \vspace{1mm}\\
	\hspace{0.15in} \textbf{Jie Li}$^1$
	\hspace{0.15in} \textbf{Hongchuan Jiang}$^1$
	\hspace{0.15in} \textbf{Zhongyue Shi}$^3$
	\hspace{0.15in} \textbf{Jun Liu}$^2$
	\hspace{0.15in} \textbf{Mulan Jin}$^{2\dag}$ \vspace{1mm}\\
	\hspace{0.1in} $^1$Department of Breast Surgery, Beijing Chao-Yang Hospital, Beijing \vspace{1mm}\\
	$^2$School of Artificial Intelligence, Beijing University of Posts and Telecommunications, Beijing \vspace{1mm}\\
	$^3$Department of Pathology, Beijing Chao-Yang Hospital, Beijing
}
\renewcommand{\shorttitle}{}

\begin{document}

\renewcommand{\thefootnote}{\fnsymbol{footnote}}
\footnotetext[1]{These authors have contributed equally to this work.}
\footnotetext[2]{Correspondence: Feng Xu (drxufeng@mail.ccmu.edu.cn), Chuang Zhu (czhu@bupt.edu.cn), Mulan Jin (kinmokuran@163.com).}

\maketitle

\begin{abstract}

\textbf{Objectives:} To develop and validate a deep learning (DL)-based primary tumor biopsy signature for predicting axillary lymph node (ALN) metastasis preoperatively in early breast cancer (EBC) patients with clinically negative ALN.

\textbf{Methods:} A total of 1,058 EBC patients with pathologically confirmed ALN status were enrolled from May 2010 to August 2020. A DL core-needle biopsy (DL-CNB) model was built on the attention-based multiple instance-learning (AMIL) framework to predict ALN status utilizing the DL features, which were extracted from the cancer areas of digitized whole-slide images (WSIs) of breast CNB specimens annotated by two pathologists. Accuracy, sensitivity, specificity, receiver operating characteristic (ROC) curves, and areas under the ROC curve (AUCs) were analyzed to evaluate our model.

\textbf{Results:} The best-performing DL-CNB model with VGG16\_BN as the feature extractor achieved an AUC of 0.816 (95\% confidence interval (CI): 0.758, 0.865) in predicting positive ALN metastasis in the independent test cohort. Furthermore, our model incorporating the clinical data, which was called DL-CNB+C, yielded the best accuracy of 0.831 (95\%CI: 0.775, 0.878), especially for patients younger than 50 years (AUC: 0.918, 95\%CI: 0.825, 0.971). The interpretation of DL-CNB model showed that the top signatures most predictive of ALN metastasis were characterized by the nucleus features including density ($p$ = 0.015), circumference ($p$ = 0.009), circularity ($p$ = 0.010), and orientation ($p$ = 0.012).

\textbf{Conclusion:} Our study provides a novel DL-based biomarker on primary tumor CNB slides to predict the metastatic status of ALN preoperatively for patients with EBC. The codes and dataset are available at \url{https://github.com/bupt-ai-cz/BALNMP}.

\end{abstract}

\keywords{deep learning \and axillary lymph node metastasis \and breast cancer \and core-needle biopsy \and whole-slide images}

\input{sections/Introduction}

\input{sections/Patiens_And_Methods}

\input{sections/Results}

\input{sections/Discussion_Conclusion}

\input{sections/Others}

\bibliographystyle{unsrtnat}
\bibliography{main}  

\input{sections/Supplementary_Material}

\end{document}

%% file: sections/Introduction.tex
\section{Introduction}

Breast cancer (BC) has become the greatest threat to women’s health worldwide \citep{Siegel2019-oe}. Clinically, identification of axillary lymph node (ALN) metastasis is important for evaluating the prognosis and guiding the treatment for BC patients \citep{Ahmed2014-nr}. Sentinel lymph node biopsy (SLNB) has gradually replaced ALN dissection (ALND) to identify ALN status, especially for early BC (EBC) patients with clinically negative lymph nodes. Although SLNB had the advantage of less invasiveness than ALND, SLNB still caused some complications such as lymphedema, axillary seroma, paraesthesia, and impaired shoulder function \citep{Kootstra2008-gl,Wilke2006-vu}. Moreover, SLNB has been considered a controversial procedure, owing to the availability of radionuclide tracers and the surgeon’s experience \citep{Manca2016-na,Hindie2011-ch}. In fact, SLNB can be avoided if there are some reliable methods of preoperative prediction of ALN status for EBC patients.

Several studies intended to predict the ALN status by clinicopathological data and genetic testing score \citep{Dihge2019-th,Shiino2019-bg}. However, due to the relatively poor predictive values and high genetic testing costs, these methods are often limited. Recently, deep learning (DL) can perform high-throughput feature extraction on medical images and analyze the correlation between primary tumor features and ALN metastasis information. In a previous study, deep features extracted from conventional ultrasound and shear wave elastography (SWE) were used to predict ALN metastasis, presenting an area under the curve (AUC) of 0.796 in the test set \citep{Zheng2020-hj}. Nevertheless, SWE has not been integrated into routine clinical breast examinations in many hospitals. Another recent study demonstrated that the DL model based on diffusion-weighted imaging–magnetic resonance imaging (DWI-MRI) database of 172 patients achieved an AUC of 0.852 for preoperative prediction of ALN metastasis \citep{Luo2018-zp}, but the small sample size enrolled could not be representative.

Currently, DL has enabled rapid advances in computational pathology \citep{Campanella2019-tg,Gu2018-wi}. For example, DL methods have been applied to segment and classify glomeruli with different staining and various pathologic changes, thus achieving the automatic analysis of renal biopsies \citep{Mei2020-ig,Jiang2021-ve}; meanwhile, DL-based automatic colonoscopy tissue segmentation and classification have shown promise for colorectal cancer detection \citep{Zhu2021-pg,Feng2020-cn}; besides, the analysis of gastric carcinoma and precancerous status can also benefit from DL schemes \citep{Iizuka2020-xq,Song2020-lw}. More recently, for the ALN metastasis detection, it is reported that DL algorithms on digital lymph node pathology images achieved better diagnostic efficiency of ALN metastasis than pathologists \citep{Hu2021-at,Zhao2020-fq}. In particular, the assistance of algorithm significantly increases the sensitivity of detection for ALN micro-metastases \citep{Steiner2018-vo}. In addition to diagnosis, several previous studies indicated that deep features based on whole-slide images (WSIs) of postoperative tumor samples potentially improved the prediction performance of lymph node metastasis in a variety of cancers \citep{Zhao2020-fq,Harmon2020-ak}. So far, there is no relevant research on preoperatively predicting ALN metastasis based on WSIs of primary BC samples. In this study, we investigated a clinical data set of EBC patients treated by preoperative core-needle biopsy (CNB) to determine whether DL models based on primary tumor biopsy slides could help to refine the prediction of ALN metastasis.

\begin{figure}[htb]
\centering
\includegraphics[width=0.7\linewidth]{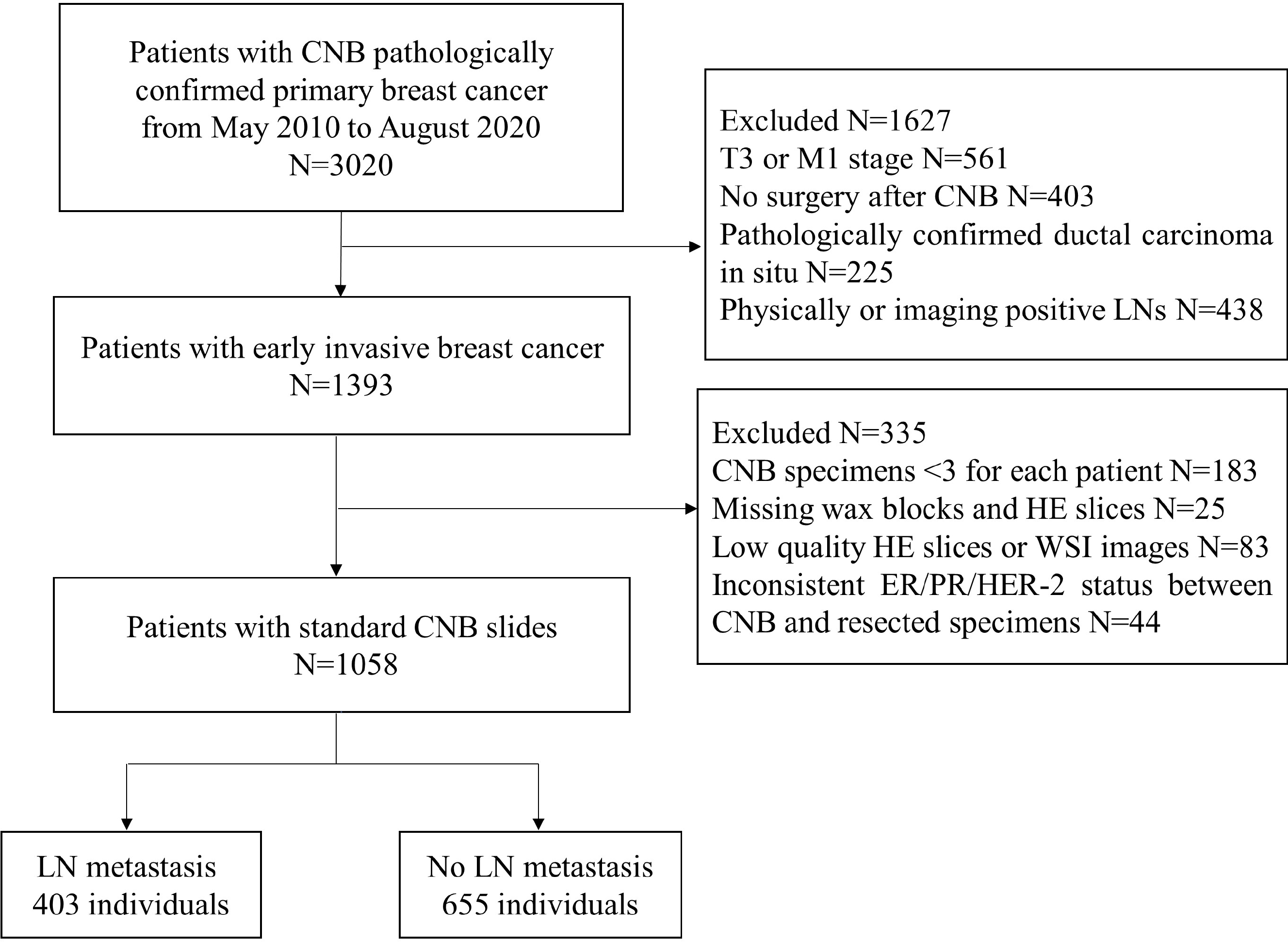}
\caption{Patient recruitment workflow.}
\label{fig:fig_1}
\end{figure}

%% file: sections/Patiens_And_Methods.tex
\section{Patients and Methods}

\subsection{Patients}

On approval by the Institutional Ethical Committees of Beijing Chaoyang Hospital affiliated to Capital Medical University, we retrospectively analyzed data from EBC patients with clinically negative ALN from May 2010 to August 2020. Written consent was obtained from all patients and their families.

The detailed inclusion criteria were as follows: 1) patients with CNB pathologically confirmed primary invasive BC; 2) patients who underwent breast surgery with SLNB or ALND; 3) baseline clinicopathological data including age, tumor size, tumor type, ER/PR/HER-2 status, and the number of ALN metastasis were comprehensive; 4) complete concordance of molecular status was found between CNB and excision specimens; 5) no history of preoperative radiotherapy and chemotherapy; and 6) adequate volume of biopsy materials with three or more cores for each patient.

The exclusion criteria included the following: 1) patients with physically positive or imaging-positive ALN; 2) missing postoperative pathology information; 3) missing wax blocks and hematoxylin and eosin (H\&E) slices; and 4) low-quality H\&E slices or WSIs. The patient recruitment workflow is shown in \figurename~\ref{fig:fig_1}.

\subsection{Deep Learning Model Development}

\begin{figure}[htb]
\centering
\includegraphics[width=0.9\linewidth]{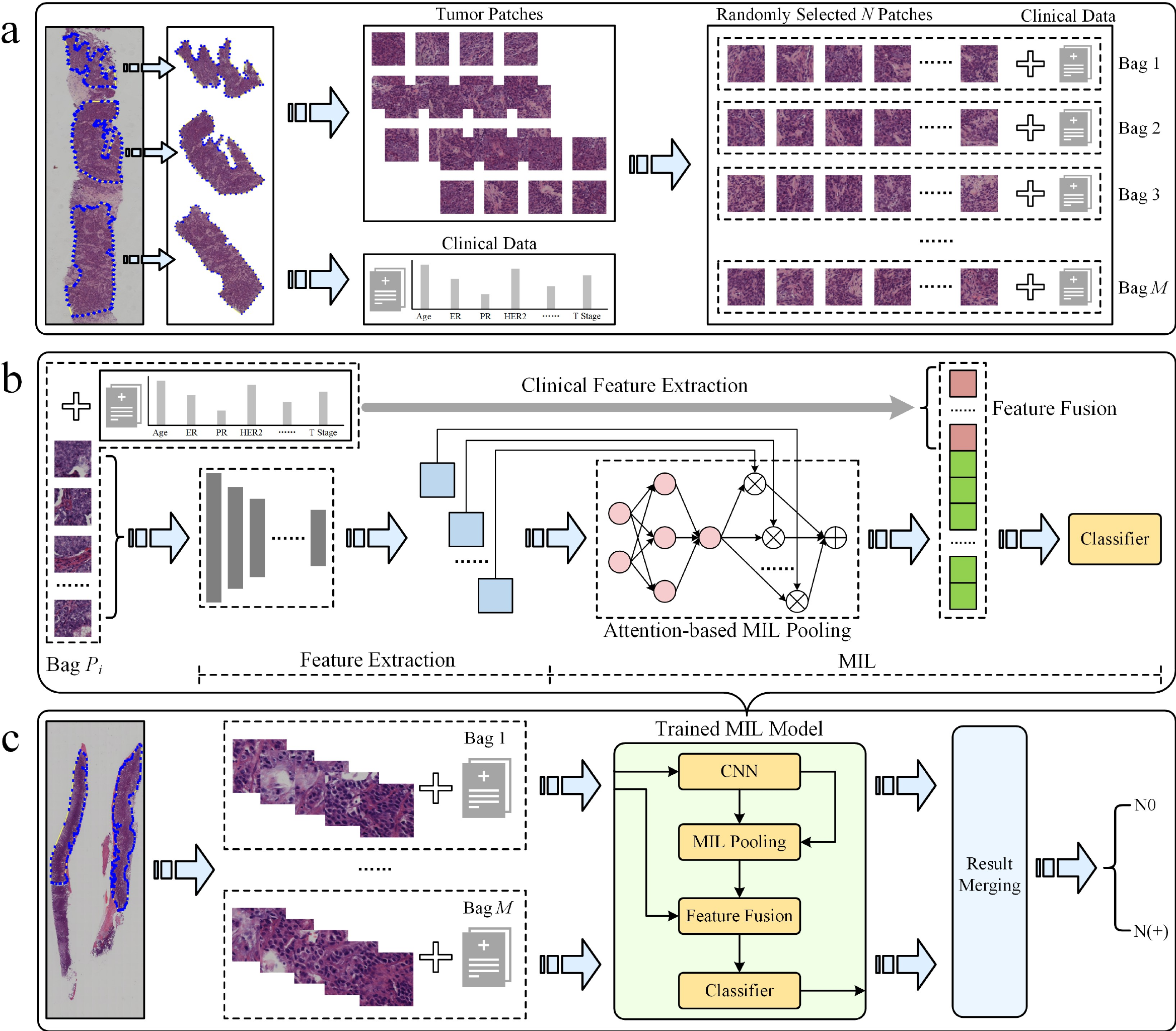}
\caption{The overall pipeline of the deep learning core-needle biopsy incorporating the clinical data (DL-CNB+C) model to predict axillary lymph node (ALN) status between N0 and N(+). (a): Multiple training bags were built based on clinical data and the cropped patches from the selected tumor regions of each core-needle biopsy (CNB) whole-slide image (WSI). (b): DL-CNB+C model training process included two phases of feature extraction and multiple-instance learning (MIL), and finally the weighted features fused with clinical features were used to predict classification probabilities and calculate the cross-entropy loss. (c): The predicted probabilities of each bag from a raw CNB WSI were merged to guide the final ALN status classification between N0 and N(+).}
\label{fig:fig_2}
\end{figure}

To avoid the inter-observer heterogeneity, all available tumor regions in each CNB slide were examined and annotated by two independent and experienced pathologists blinded to all patient-related information. A WSI was classified into positive (N(+)) or negative (N0) using the proposed DL CNB (DL-CNB) model. Our DL-CNB model was constructed with the attention-based multiple-instance learning (MIL) approach \citep{Ilse2018-mn}. In MIL, each training sample was called a bag, which consisted of multiple instances \citep{Das2018-po,Sudharshan2019-vy,Couture2018-hx} (each instance corresponds to an image patch of size 256$\times$256 pixels). Different from the general fully supervised problem where each sample had a label, only the label of bags was available in MIL, and the goal of MIL was to predict the bag label by considering all included instances comprehensively. The whole algorithm pipeline comprised the following five steps:

(1) Training data preparation (\figurename~\ref{fig:fig_2}a). For each raw WSI, amounts of non-overlapping square patches were first cropped from the selected tumor regions. Then each WSI could be represented as a bag with $N$ randomly selected patches. To increase the training samples, $M$ bags were built for each WSI. All $M$ bags were labeled as positive if the slide is an ALN metastasis case, and vice versa. Note that we could add the clinical information of the slide to all the $M$ constructed bags to involve more useful information for predicting, and in this situation, the developed model was called DL-CNB+C.

(2) Feature extraction (left part of \figurename~\ref{fig:fig_2}b). $N$ feature vectors were extracted for the $N$ image instances in each bag by using a convolutional neural network (CNN) model. The performances of AlexNet \citep{Krizhevsky2012-xe}, VGG16 \citep{Simonyan2015-ri} with batch norm (VGG16\_BN), ResNet50 \citep{He2016-lq}, DenseNet121 \citep{Huang2017-di}, and Inception-v3 \citep{Szegedy2016-cv} were compared to find the best feature extractor. At this stage, the clinical data were also preprocessed for feature extraction. Concretely, the numerical properties in clinical data were standardizing by removing the mean and scaling to unit variance, thus eliminating the effect of data range and scale; furthermore, considering that there was no natural ordinal relationship between different values of the category attributes, the categorical properties in clinical data were encoded as the one-hot vectors, which could express different values equally.

(3) MIL (right part of \figurename~\ref{fig:fig_2}b).The extracted $N$ feature vectors of image instances were first processed by the max-pooling \citep{Feng2017-hk,Pinheiro2015-ud,Zhu2017-tc} and reshaping and then were passed to a two-layer fully connected (FC) layer. The $N$ weight factors for the instances in the bag were thus obtained and then were further multiplied to the original feature vectors \citep{Ilse2018-mn} to adaptively adjust the effect of instance features. Finally, the weighted image feature vectors and the clinical features were fused by concatenation; due to the large difference of dimensions between image features and clinical features, the clinical features were copied 10 times for expansion. Then, the fused features were fed into the classifier, and the outputs and the ground truth labels were used to calculate the cross-entropy loss.

(4) Model training and testing. We randomly divided the WSIs into training cohort and independent test cohort with the ratio of 4:1 and randomly selected 25\% of the training cohort as the validation cohort. We used Adam optimizer with learning rate 1e-4 to update the model parameters and weight decay 1e-3 for regularization. In the training phase, we used the cosine annealing warm restarts strategy to adjust the learning rate \citep{Loshchilov2017-mo}. In the testing phase, the ALN status is predicted by aggregating the model outputs of all bags from the same slide (\figurename~\ref{fig:fig_2}c).

The deep learning models are available at \url{https://github.com/bupt-ai-cz/BALNMP}.

\subsection{Visualization of Salient Regions From Deep Learning Core-Needle Biopsy Model}

We visualized the important regions that were more associated with metastatic status. After the processing of attention-based MIL pooling, the weights of different patches can be obtained, and the corresponding feature maps were then weighted together in the following FC layers to conduct ALN status prediction. With the attention weights, we created a heat map to visualize the important salient regions in each WSI.

\subsection{Interpretability of Deep Learning Core-Needle Biopsy Model With Nucleus Features}

\begin{figure}[htb]
\centering
\includegraphics[width=1.0\linewidth]{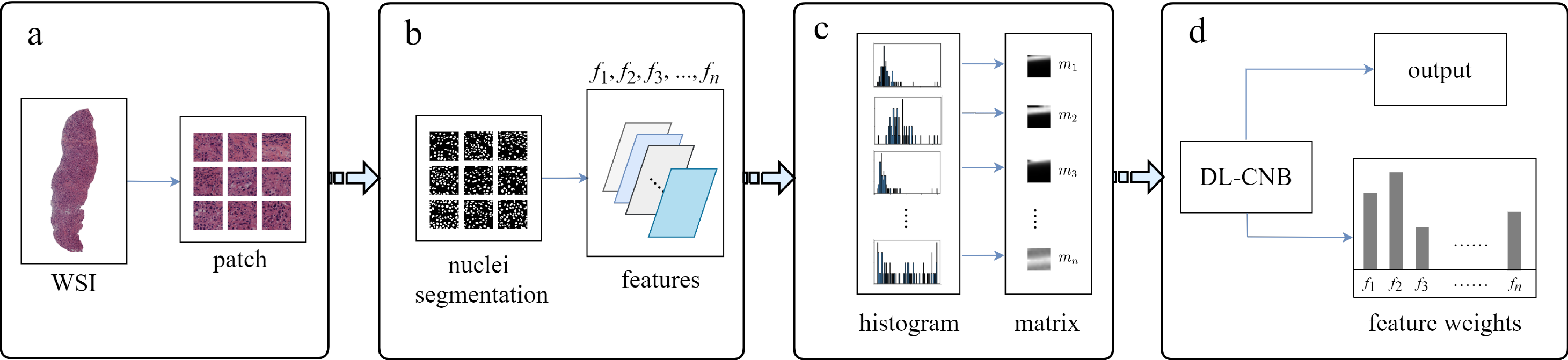}
\caption{Overview on interpretability methods of deep learning core-needle biopsy (DL-CNB) model based on nucleus morphometric features. (a): The selected tumor regions of each whole-slide image (WSI) was cropped into patches. (b): For each patch, we processed nucleus segmentation (a weakly supervised segmentation framework was applied to obtain the nucleus), defined multiple nucleus morphometric features (such as major axis, minor axis, area, orientation, circumference, density, circularity, and rectangularity, which are denoted as $f_1, f_2, f_3, ..., f_n$), and extracted $n$ feature parameters correspondingly. (c): All $n$ kinds of feature parameters from a WSI were quantized into $n$ distribution histograms and saved to $n$ feature matrices ($m_1, m_2, m_3, ..., m_n$). (d) The matrices from a WSI were considered as instances of a bag and served as the input of DL-CNB model; the re-trained DL-CNB model could generate scores of features (instances) in the bag, which represented the weight of each feature in pathological diagnosis.}
\label{fig:fig_3}
\end{figure}

Interpretability of DL-CNB model with nucleus features was performed to study the contribution of different nucleus morphological characteristics in the prediction of lymph node metastasis \citep{Mueller2016-fl,Radhakrishnan2017-hm}. Multiple specially designed nucleus features were firstly extracted for each WSI, and these features together formed a training bag. With the constructed feature bags, the proposed DL-CNB model was re-trained. The weights of different features (instances) can be obtained based on the attention-based MIL pooling, and thus the contribution of different features was yielded. The specific process is described in \figurename~\ref{fig:fig_3}.

\subsection{Statistical Analysis}

The logistic regression was used to predict ALN status by clinical data only model. The clinical difference of N0 and N(+) was compared by using the Mann–Whitney U test and chi-square test. The AUCs of different methods were compared by using Delong et al. \citep{DeLong1988-hn}. The other measurements like accuracy (ACC), sensitivity (SENS), specificity (SPEC), positive predictive value (PPV), and negative predictive value (NPV) were also used to estimate the model performance. All the statistics were two-sided, and a $p$-value less than 0.05 was considered statistically significant. All statistical analyses were performed by MedCalc software (V 19.6.1; 2020 MedCalc Software bvba, Mariakerke, Belgium), Python 3.7, and SPSS 24.0 (IBM, Armonk, NY, USA).

%% file: sections/Results.tex
\section{Results}

\subsection{Clinical Characteristics}

A total of 1,058 patients with EBC were enrolled for analysis. Among them, 957 (90.5\%) patients had invasive ductal carcinomas, 25 (2.4\%) patients had invasive lobular carcinomas, and 76 (7.1\%) patients had other types. There were 840 patients in the training cohort and 218 patients in the independent test cohort after all WSIs were randomly divided by using N0 as the negative reference standard and others as the positive. The average patient age was 57.6 years (range, 26-90 years) for the training and validation sets and 56.7 years (range, 22-87 years) for the test set. The mean ultrasound tumor size was 2.23 cm (range, 0.5-4.5 cm). A total of 556 patients (52.6\%) had T1 tumors, while 502 patients (47.4\%) had T2 tumors. According to the results of SLNB or ALND, positive lymph nodes were found in 403 patients. Among them, 210 patients (52.1\%) had one or two positive lymph nodes (N+(1-2)), and 193 patients (47.9\%) had three or more positive lymph nodes (N+($\geq$ 3)). As shown in \tablename~\ref{tab:table_1}, there was no significant difference between the detailed characteristics of the training and independent test cohorts (all $p \ge$ 0.05).

\subsection{Convolutional Neural Network Model Selection}

The detailed results are summarized in supplementary \tablename~\ref{tab:table_s_1}. Based on the overall analysis, VGG16\_BN model pre-trained on ImageNet \citep{Deng2009-im} provided the best performance in the validation cohort and the independent test cohort (AUC: 0.808, 0.816), compared with AlexNet (AUC: 0.764, 0.780), ResNet50 (AUC: 0.644, 0.607), DenseNet121 (AUC: 0.714, 0.739), and Inception-v3 (AUC: 0.753, 0.762). Furthermore, considering other metrics, VGG16\_BN achieved the best ACC, SPEC, and PPV in the independent test cohort. VGG16\_BN consisted of (convolution layer, batch normalization layer, and Rectified Linear Unit (ReLU)) as the basic block where ReLU played a role of activation function to provide the non-linear capability; and max-pooling layers were inserted between basic blocks for down-sampling; besides, there was an adaptive average pooling layer at the end of VGG16\_BN for obtaining features with a fixed size. The details of VGG16\_BN are described in supplementary \tablename~\ref{tab:table_s_2}.

\subsection{Predictive Value of Deep Learning Core-Needle Biopsy Incorporating the Clinical Data Model Between N0 and N(+)}

In the training cohort, DL-CNB+C achieved an AUC of 0.878, while DL-CNB and classification by clinical data only model achieved AUCs of 0.901 and 0.661, respectively. And in the validation cohort, the DL-CNB+C model achieved an AUC of 0.823, which was higher than an AUC of 0.808 obtained by DL-CNB only and an AUC of 0.709 obtained by classification by clinical data.

\begin{figure}[htb]
\centering
\includegraphics[width=0.6\linewidth]{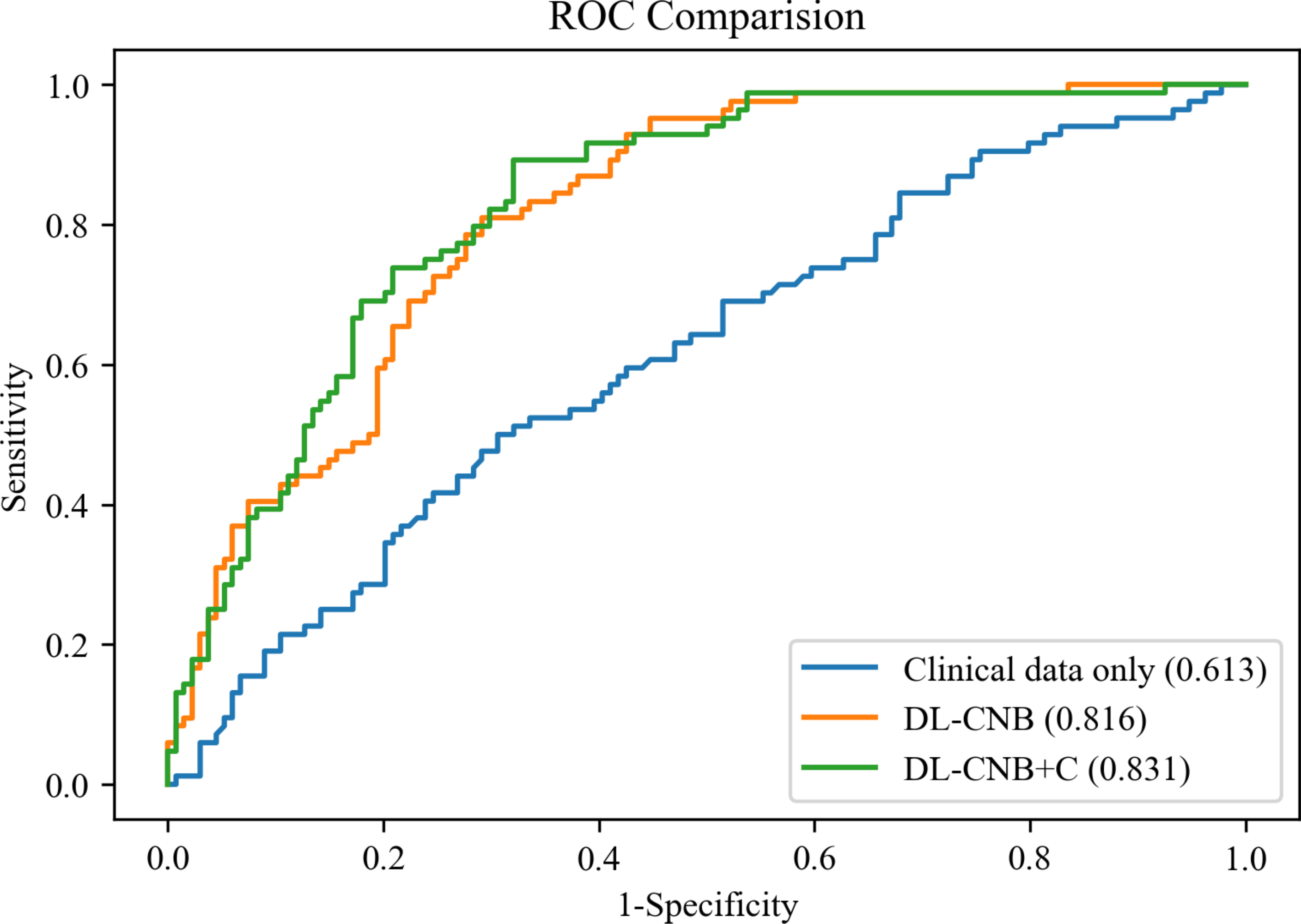}
\caption{Comparison of receiver operating characteristic (ROC) curves between different models for predicting disease-free axilla (N0) and heavy metastatic burden of axillary disease (N(+)). Numbers in parentheses are areas under the receiver operating characteristic curve (AUCs).}
\label{fig:fig_4}
\end{figure}

In the independent test cohort, the DL-CNB+C model still achieved the highest AUC of 0.831, which was better than the AUC of DL-CNB only (AUC: 0.816, $p$ = 0.453) and classification by clinical data only (AUC: 0.613, $p \le$ 0.0001). The ACC, SENS, and NPV of DL-CNB+C were also better than those of other methods. The detailed statistical results are summarized in \tablename~\ref{tab:table_2}, and its corresponding receiver operating characteristics (ROCs) are shown in \figurename~\ref{fig:fig_4}.

We further divided N(+) into low metastatic potential (N$_+$(1-2)) and high metastatic potential (N$_+$($\geq$3)) according to the number of ALN metastasis. Adopting N0 as the negative reference standard, the combined model showed better discriminating ability between N0 and N$_+$(1-2) (AUC: 0.878) and between N0 and N$_+$($\geq$3) (AUC: 0.838).

The detailed statistical results are summarized in supplementary \tablename~\ref{tab:table_s_3} and supplementary \tablename~\ref{tab:table_s_4}, and the corresponding ROCs are shown in supplementary \figurename~\ref{fig:fig_s_1} and supplementary \figurename~\ref{fig:fig_s_2}.

\begin{table}[htb]
\centering
\caption{Patient and tumor characteristics.}
\label{tab:table_1}
\resizebox{0.8\linewidth}{!}{
\renewcommand\arraystretch{1.3}
\begin{threeparttable}
\begin{tabular}{llllll}
\toprule
Characteristics             &                            & All patients  & Training  & Test          & $p$                      \\
\midrule
Number                      &                            & 1058          & 840 (80\%)    & 218 (20\%)    &                        \\
Age, mean $\pm$ SD, years      &                            & 57.58$\pm$12.523 & 57.80$\pm$12.481 & 56.72$\pm$12.674 & 0.344                  \\
Tumor size, mean $\pm$ SD, cm  &                            & 2.234$\pm$0.8623 & 2.228$\pm$0.8516 & 2.256$\pm$0.9040 & 0.898                  \\
Number of LNM, mean $\pm$ SD     &                            & 1.20$\pm$2.081   & 1.20$\pm$2.095   & 1.20$\pm$2.033   & 0.847                  \\
Tumor type                  & Invasive ductal carcinoma  & 957          & 760 (90.5\%)  & 197 (90.4\%)  & 0.812 \\
                            & Invasive lobular carcinoma & 25           & 20 (2.4\%)    & 5 (2.3\%)    &       \\
                            & Other types                & 76           & 60 (78.9\%)    & 16 (21.1\%)    &       \\
T stage    & T1                         & 556           & 435 (51.8\%)  & 121 (55.5\%)  & 0.327 \\
                            & T2                         & 502           & 405 (48.2\%)  & 97 (44.5\%)   &        \\
ER         & Positive                   & 831           & 665 (79.2\%)  & 166 (76.1\%)  &  0.333 \\
           & Negative                   & 227           & 175 (20.8\%)  & 52 (23.9\%)   &         \\
PR         & Positive                   & 790           & 633 (75.4\%)  & 157 (72.0\%)  & 0.312 \\
           & Negative                   & 268           & 207 (24.6\%)  & 61 (28.0\%)   &       \\
HER2       & Positive                   & 277           & 217 (25.8\%)  & 60 (27.5\%)   & 0.613 \\
           & Negative                   & 781           & 623 (74.2\%)  & 158 (72.5\%)  &        \\
Molecular subtype   & Luminal A	& 288	& 223 (26.5\%)  & 65 (29.8\%) & 0.556 \\
                    & Luminal B	& 372	& 304 (36.2\%)	& 68 (31.2\%) & \\
                    & Triple negative &	125	& 99 (11.8\%) &	26 (11.9\%) & \\
                    & HER2(+) &	273 &	214 (25.5\%) &	59 (27.1\%) & \\
LNM        & Yes                        & 403           & 521 (62.0\%)  & 134 (61.5\%)  & 0.880 \\
           & No                         & 655           & 319 (38.0\%)  & 84 (38.5\%)   &       \\
\bottomrule
\end{tabular}
\begin{tablenotes}
\footnotesize
\item Qualitative variables are in $n$ (\%), and quantitative variables are in mean $\pm$ SD, when appropriate.
\item SD, standard deviation; ER, estrogen receptor; PR, progesterone receptor; HER-2, human epidermal growth factor receptor-2; LNM, lymph node metastasis.
\end{tablenotes}
\end{threeparttable}
}
\end{table}

\begin{table}[htb]
\centering
\caption{The performance in prediction of ALN status (N0 vs. N(+)).}
\label{tab:table_2}
\resizebox{1\linewidth}{!}{
\renewcommand\arraystretch{1.3}
\begin{threeparttable}
\begin{tabular}{llllllll}
\toprule
Methods &  & AUC & ACC (\%) & SENS (\%) & SPEC (\%) & PPV (\%) & NPV (\%) \\
\midrule
\multirow{3}{*}{Clinical data only} & T & 0.661   {[}0.622, 0.698{]} & 64.13 {[}60.24, 67.88{]} & 64.58 {[}58.17, 70.63{]} & 63.85 {[}58.86, 68.62{]} & 52.36 {[}48.32, 56.38{]} & 74.55 {[}70.85, 77.92{]} \\
 & V & 0.709   {[}0.643, 0.770{]} & 67.62 {[}60.84, 73.90{]} & 65.82 {[}54.29, 76.13{]} & 68.70 {[}60.02, 76.52{]} & 55.91 {[}48.46, 63.11{]} & 76.92 {[}70.62, 82.22{]} \\
 & I-T & 0.613$^{a, b}$ {[}0.545, 0.678{]} & 61.93 {[}55.12, 68.40{]} & 50.00 {[}38.89, 61.11{]} & 69.40 {[}60.86, 77.07{]} & 50.60 {[}42.34, 58.83{]} & 68.89 {[}63.49, 73.82{]} \\
\midrule
\multirow{3}{*}{DL-CNB model} & T & 0.901   {[}0.875, 0.923{]} & 80.32 {[}76.99, 83.35{]} & 94.17 {[}90.41, 96.77{]} & 71.79 {[}67.05, 76.21{]} & 67.26 {[}63.61, 70.71{]} & 95.24 {[}92.30, 97.09{]} \\
 & V & 0.808   {[}0.748, 0.859{]} & 72.86 {[}66.31, 78.75{]} & 77.22 {[}66.40, 85.90{]} & 70.23 {[}61.62, 77.90{]} & 61.00 {[}53.95, 67.62{]} & 83.64 {[}77.04, 88.62{]} \\
 & I-T & 0.816$^{c}$   {[}0.758, 0.865{]} & 74.77 {[}68.46, 80.39{]} & 80.95 {[}70.92, 88.70{]} & 70.90 {[}62.43, 78.42{]} & 63.55 {[}56.76, 69.84{]} & 85.59 {[}79.04, 90.34{]} \\
\midrule
\multirow{3}{*}{DL-CNB+C model} & T & 0.878 {[}0.622, 0.698{]} & 76.51 {[}73.00, 79.77{]} & 93.33 {[}89.40, 96.14{]} & 66.15 {[}61.22, 70.84{]} & 62.92 {[}59.53, 66.19{]} & 94.16 {[}90.90, 96.30{]} \\
 & V & 0.823 {[}0.765, 0.872{]} & 75.71 {[}69.34, 81.35{]} & 74.68 {[}63.64, 83.80{]} & 76.34 {[}68.12, 83.32{]} & 65.56 {[}57.69, 72.65{]} & 83.33 {[}77.19, 88.08{]} \\
 & I-T & 0.831 {[}0.775, 0.878{]} & 75.69 {[}69.44, 81.23{]} & 89.29 {[}80.63, 94.98{]} & 67.16 {[}58.53, 75.03{]} & 63.03 {[}56.96, 68.71{]} & 90.91 {[}84.21, 94.94{]} \\
\bottomrule
\end{tabular}
\begin{tablenotes}
\footnotesize
\item 95\% confidence intervals are included in brackets.
\item AUC, area under the receiver operating characteristic curve; ACC, accuracy; SENS, sensitivity; SPEC, specificity; PPV, positive predictive value; NPV, negative predictive value.
\item T, training cohort ($n$ = 630); V, validation cohort ($n$ = 210); I–T, independent test cohort ($n$ = 218).
\item ALN, axillary lymph node; DL-CNB+C, deep learning core-needle biopsy incorporating the clinical data.
\item $^{a}$ Indicates $p$ < 0.0001, Delong et al. in comparison with DL-CNB model in independent test cohort.
\item $^{b}$ Indicates $p$ < 0.0001, Delong et al. in comparison with DL-CNB+C model in independent test cohort.
\item $^{c}$ Indicates $p$ = 0.4532, Delong et al. in comparison with DL-CNB+C model in independent test cohort.
\end{tablenotes}
\end{threeparttable}
}
\end{table}

\subsection{Predictive Value of Deep Learning Core-Needle Biopsy Incorporating the Clinical Data Model Among N0, N$_+$(1-2) and N$_+$($\geq$3)}

\begin{figure}[htb]
\centering
\includegraphics[width=0.6\linewidth]{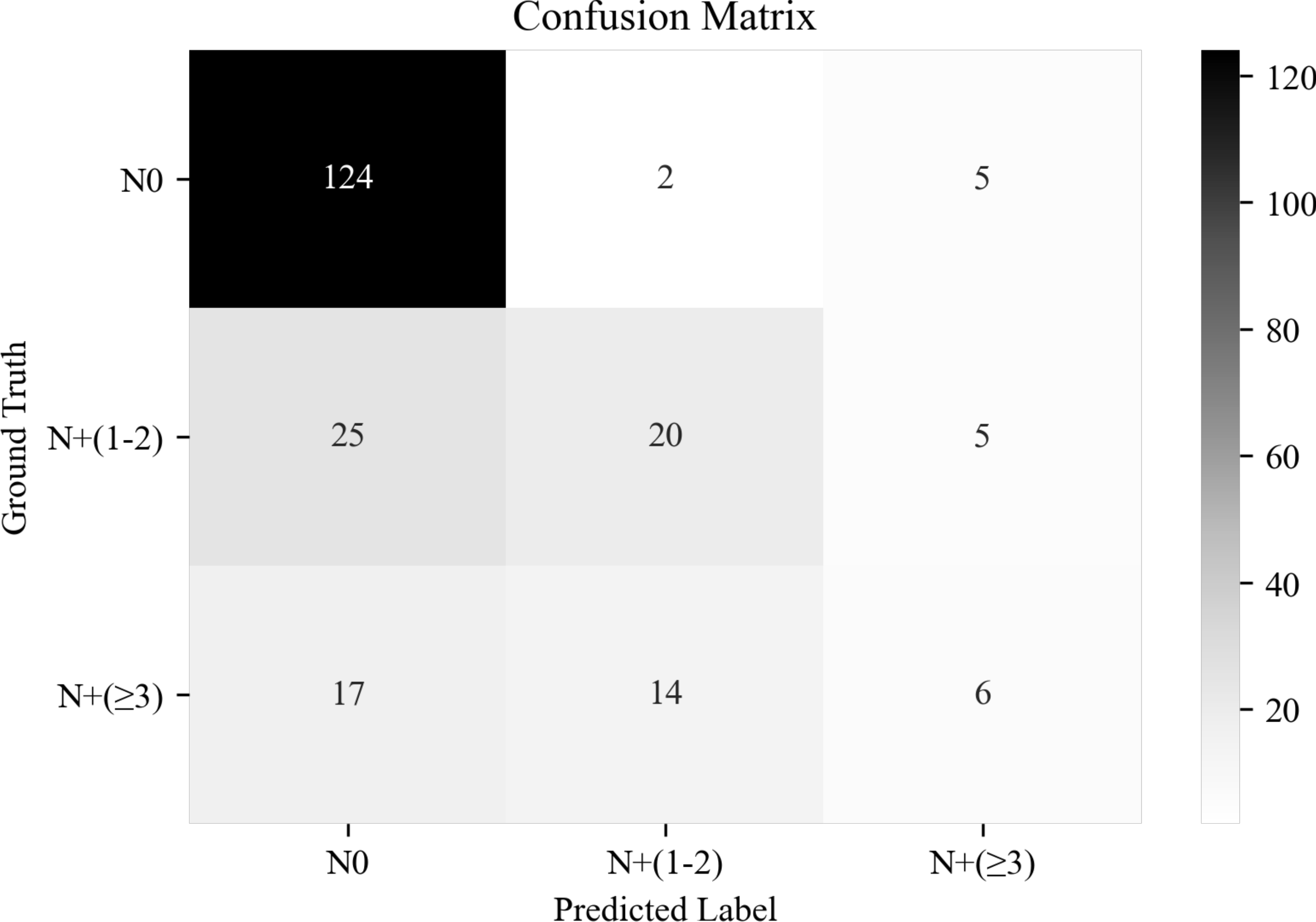}
\caption{The confusion matrix of predicting axillary lymph node (ALN) status between disease-free axilla (N0), low metastatic burden of axillary disease (N$_+$(1-2)), and heavy metastatic burden of axillary disease (N$_+$($\geq$3)).}
\label{fig:fig_5}
\end{figure}

The overall AUC of multi-classification in the independent test cohort based on DL-CNB+C model was 0.791; there existed the highest precision and recall of 0.747 and 0.947, respectively, in N0; there existed the precision and recall of 0.556 and 0.400 in N$_+$(1-2); and there existed the precision and recall of 0.375 and 0.162 in N$_+$($\geq$3). The confusion matrix under the classification threshold of 0.5 is shown in \figurename~\ref{fig:fig_5}. According to the results, the model performed well in differentiating the N0 group while showing poor diagnostic efficacy in the other two groups.

\subsection{Subgroup Analysis of Deep Learning Core-Needle Biopsy Incorporating the Clinical Data Model}

Furthermore, we analyzed the measurement results of the different subgroups in the independent test cohort of predicting ALN status between N0 and N(+) by the DL-CNB+C model. The detailed statistical results are summarized in supplementary \tablename~\ref{tab:table_s_5}. In the independent test cohort, compared with an AUC of 0.794 (95\%CI: 0.720, 0.855) in the subgroup of age $>$ 50, there existed better performance in the subgroup of age $\leq$ 50 with an AUC of 0.918 (95\%CI: 0.825, 0.971, $p$ = 0.015). There were no significant differences regarding other subgroups of ER(+) vs. ER(-) ($p$ = 0.125), PR(+) vs. PR(-) ($p$ = 0.659), HER-2(+) vs. HER-2(-) ($p$ = 0.524), and T1 vs. T2 stage ($p$ = 0.743) between N0 and N(+).

\subsection{Interpretability of Deep Learning Core-Needle Biopsy Model}

\begin{figure}[htb]
\centering
\includegraphics[width=0.65\linewidth]{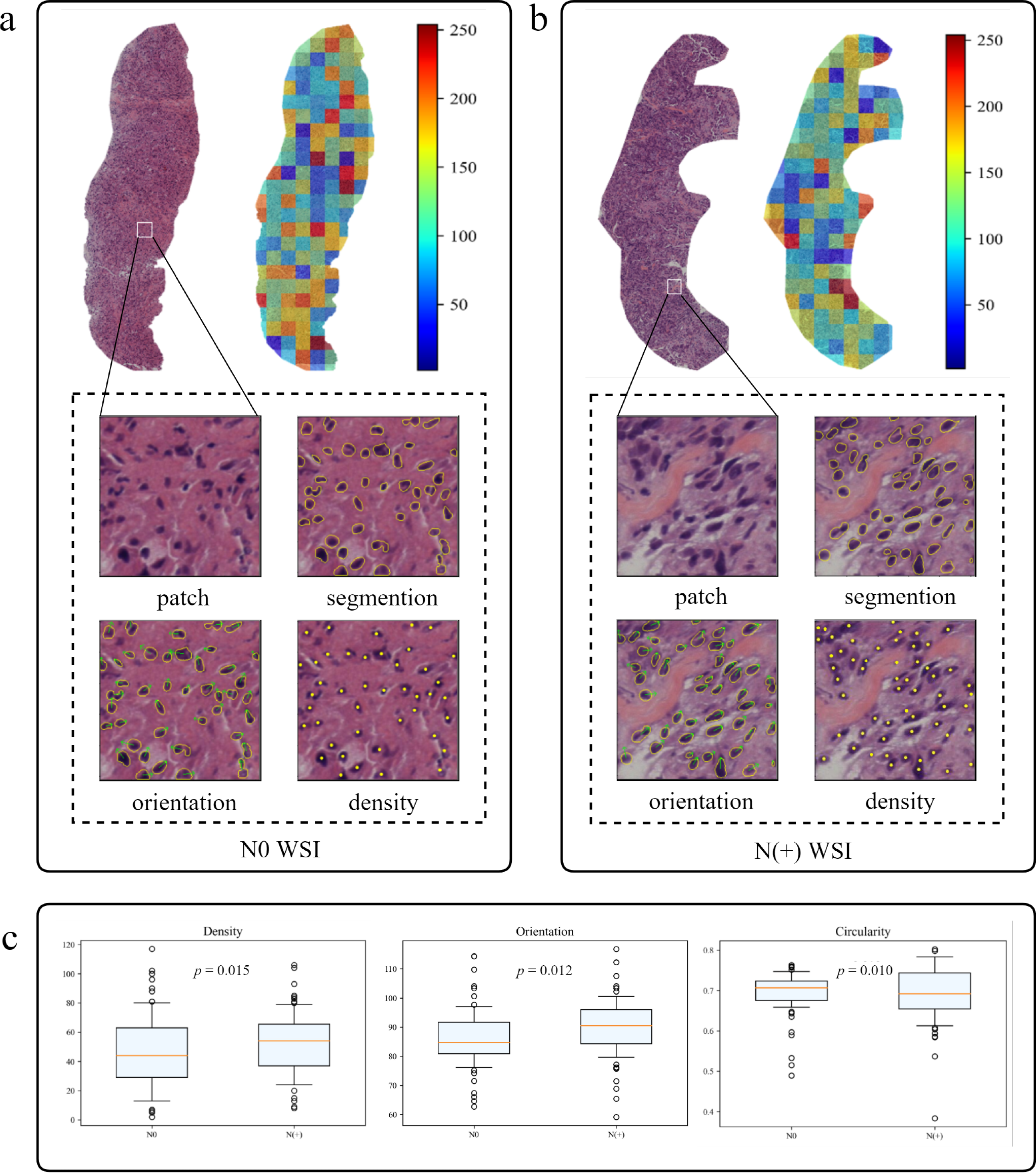}
\caption{The interpretability of the deep learning core-needle biopsy (DL-CNB) model of two patients. (a, b) The heat maps and nuclear segmentation from core-needle biopsy (CNB) whole-slide images (WSIs) of the N0 and the N(+) separately, and the red regions show greater contribution to the final classification. (c) The statistical analysis of three nuclear characteristics most relevant to diagnosis of all patients.}
\label{fig:fig_6}
\end{figure}

To investigate the interpretability of the DL-CNB, we conducted two studies for digging the correlation factors of ALN status prediction. In the first study, we adopted the attention-based MIL pooling to find the important regions that contributing to the prediction. The heat map in \figurename~\ref{fig:fig_6}a highlights the red patches as the important regions. Although the obtained important areas can provide some clues to the diagnosis of DL-CNB model, it is not clear that the model makes decisions based on what features of the tumor area.

In the second study, we specially designed and extracted multiple nucleus features for each WSI. The weights of different features were then obtained based on the same attention-based MIL pooling in our DL-CNB. The weights highlighted the nucleus features that were most relevant to the ALN status prediction of each WSI. We found that the WSI of N(+) group had higher nuclear density ($p$ = 0.015) and orientation ($p$ = 0.012) but lower circumference ($p$ = 0.009), circularity ($p$ = 0.010), and area ($p$ = 0.024) compared with N0 group (\figurename~\ref{fig:fig_6}b and \figurename~\ref{fig:fig_6}c). There were no significant differences in other nucleus features including major axis ($p$ = 0.083), minor axis ($p$ = 0.065), and rectangularity ($p$ = 0.149) between N0 and N(+).

%% file: sections/Discussion_Conclusion.tex
\section{Discussion}

In most previous studies, DL signatures of ALN metastases were based on medical images such as ultrasound, CT, and MRI \citep{Luo2018-zp,Zhou2020-bx,Yang2020-qi}. However, since many patients had undergone CNB at the time of imaging examination, and the reactive changes such as needle path in the tumor would result in the predictive inaccuracy of imaging information. This study focused on preoperative CNB WSI, which also played an important role in BC management and has been increasingly performed in clinical practice. Preoperative CNB can provide not only the histopathological diagnosis of BC but also the molecular status including ER/PR/HER-2 status, which is associated with ALN metastasis \citep{Calhoun2014-ja}. Otherwise, the morphological features of tumor cells can be visualized on CNB WSI. Therefore, primary tumor biopsy WSI as a complementary imaging tool has the potential for ALN metastasis prediction. To the best of our knowledge, this is the first study to apply the DL-based histopathological features extracted from primary tumor WSIs for ALN prediction analysis.

Here, the best-performing DL-CNB model yielded satisfactory predictions with an AUC of 0.816, a SENS of 81.0\%, and a SPEC of 70.9\% on the test set, which had superior predictive capability as compared with clinical data alone. Furthermore, unlike other combined models incorporating clinical data \citep{Dihge2019-th,Zheng2020-hj}, the DL-CNB+C model slightly improved the ACC to 0.831, which showed that our results were mainly derived from the contribution of DL-CNB model. In addition, during the subgroup analysis stratified by patient’s age, our DL-CNB+C model achieved an AUC of 0.918 for patients younger than 50 years, indicating that age was the critical factor in predicting ALN status. Regarding the number of ALN metastasis, the DL-CNB+C model showed better discriminating ability between N0 an N+(1-2), and between N0 and N+($\geq$3). However, the unfavorable discriminating ability was found between N+(1-2) and N+($\geq$3). This was consistent with the study of Zheng et al. \citep{Zheng2020-hj}, who also reported poor efficacy between N+(1-2) and N+($\geq$3), utilizing the DL radiomics model. In the future, further exploration of ALN staging prediction is needed.

Indeed, computer-assisted histopathological analysis can provide a more practical and objective output\cite{Acs2020-gn}. For example, different molecular subtypes \citep{Jaber2020-ct} and Oncotype DX risk score \citep{Whitney2018-ju} occurring in BC could be directly predicted from the H\&E slides. On the one hand, our DL model can provide significant information for risk stratification and axillary staging, thereby avoiding axillary surgery and reducing the complication and hospitalization costs. On the other hand, our results also highlight the development of algorithms based on artificial intelligence, which will reduce the labor intensity of pathologists. Similar approaches may be used to the pathology of other organs.

In our study, we are first to quantitatively assess the role of nuclear disorder in predicting ALN metastasis in BC. Our finding is consistent with several recent studies that demonstrate the powerful predictive effect of nuclear disorder on patient survival \citep{Lu2018-vq,Lee2017-nj}. Interestingly, the top predictive signatures that distinguished N0 from N(+) were characterized by the nucleus features including density, circumference, circularity, and orientation. We found that the WSI of N(+) had higher nuclear density and polarity but lower circularity, which was understandable since in the tumors with ALN metastasis, tumor cells became poorly differentiated as a result of rapid cell growth, encouraging the nuclei in these structures to form highly clustered and consistently metastatic patterns. Our results showed that nuanced patterns of nucleus density and orientation of tumor cells are important determinants of ALN metastasis.

There are some limitations in our study. First, the selection of regions of interest within each CNB slide required pathologist guidance. Future studies will explore more advanced methods for automatic segmentation of tumor regions. Second, this is a retrospective study, and prospective validation of our model in a large multicenter cohort of EBC patients is necessary to assess the clinical applicability of the biomarker. Third, recent evidence indicated that a set of features related to tumor-infiltrating lymphocytes (TILs) was found to be associated with positive LNs in bladder cancer \citep{Harmon2020-ak}. However, due to few TILs on breast CNB slides, we only selected sufficient tumor cells for the identification of salient regions rather than whole slides. Finally, we only chose H\&E stained images of CNB samples. The clinical utility of immunochemical stained images remains to be established as an interesting attempt.

\section{Conclusion}

In brief, we demonstrated that a novel DL-based biomarker on primary tumor CNB slides predicted ALN metastasis preoperatively for EBC patients with clinically negative ALN, especially for younger patients. Our methods could help to avoid unnecessary axillary surgery based on the widely collected H\&E-stained histopathology slides, thereby contributing to precision oncology treatment.

%% file: sections/Others.tex
\section*{Data availability statement}

The original contributions presented in the study are included in the supplementary material. Further inquiries can be directed to the corresponding authors.

\section*{Ethics statement}

Written informed consent was obtained from the individual(s) for the publication of any potentially identifiable images or data included in this article.

\section*{Author contributions}

FX, CZ, JiL, YW, and MJ designed the study. CZ, WT, YZ, and JiL trained the model. FX, YW, ZS, JuL,and HJ collected the data. FX, WT, YZ, CZ, YW, MJ, and JuL analyzed and interpreted the data. FX, CZ, WT, YZ, and MJ prepared the manuscript. All authors contributed to the article and approved the submitted version.

\section*{Funding}

The work was supported by National Natural Science Foundation of China [No. 8197101438].

%% file: sections/Supplementary_Material.tex
\appendix

\section*{Supplementary material}

\renewcommand{\theHtable}{A.Tab.\arabic{table}}
\renewcommand{\theHfigure}{A.Abb.\arabic{figure}}
\setcounter{table}{0}
\setcounter{figure}{0}

\begin{table}[htb]
\centering
\caption{The performance comparison of different base models in prediction of ALN status (N0 vs. N(+)).}
\label{tab:table_s_1}
\resizebox{1\linewidth}{!}{
\renewcommand\arraystretch{1.3}
\begin{threeparttable}
\begin{tabular}{llllllll}
\toprule
Base   models &  & AUC & ACC   (\%) & SENS   (\%) & SPEC   (\%) & PPV   (\%) & NPV   (\%) \\
\midrule
\multirow{3}{*}{AlexNet} & T & 0.909   {[}0.884, 0.930{]} & 82.70   {[}79.51, 85.57{]} & 88.33   {[}83.58, 92.11{]} & 79.23   {[}74.86, 83.15{]} & 72.35   {[}68.20, 76.16{]} & 91.69   {[}88.59, 94.01{]} \\
 & V & 0.764   {[}0.700, 0.819{]} & 65.71   {[}58.87, 72.11{]} & 89.87   {[}81.02, 95.53{]} & 51.15   {[}42.26, 59.97{]} & 52.59   {[}47.84, 57.30{]} & 89.33   {[}80.96, 94.28{]} \\
 & I-T & 0.780   {[}0.719, 0.833{]} & 73.39   {[}67.01, 79.13{]} & 83.33   {[}73.62, 90.58{]} & 67.16   {[}58.53, 75.03{]} & 61.40   {[}55.08, 67.36{]} & 86.54   {[}79.71, 91.32{]} \\
\midrule
\multirow{3}{*}{ResNet50} & T & 0.912   {[}0.887, 0.933{]} & 85.71   {[}82.74, 88.35{]} & 85.83   {[}80.77, 89.99{]} & 85.64   {[}81.76, 88.97{]} & 78.63   {[}74.17, 82.50{]} & 90.76   {[}87.77, 93.08{]} \\
 & V & 0.644   {[}0.575, 0.709{]} & 59.52   {[}52.55, 66.22{]} & 70.89   {[}59.58, 80.57{]} & 52.67   {[}43.77, 61.45{]} & 47.46   {[}41.80, 53.19{]} & 75.00   {[}67.22, 81.44{]} \\
 & I-T & 0.607   {[}0.539, 0.673{]} & 58.72   {[}51.87, 65.32{]} & 66.67   {[}55.54, 76.58{]} & 53.73   {[}44.92, 62.38{]} & 47.46   {[}41.61, 53.37{]} & 72.00   {[}64.65, 78.33{]} \\
\midrule
\multirow{3}{*}{DenseNet121} & T & 0.967   {[}0.949, 0.979{]} & 89.84   {[}87.21, 92.09{]} & 95.83   {[}92.47, 97.98{]} & 86.15   {[}82.32, 89.42{]} & 80.99   {[}76.85, 84.53{]} & 97.11   {[}94.82, 98.41{]} \\
 & V & 0.714   {[}0.648, 0.774{]} & 68.57   {[}61.82, 74.79{]} & 73.42   {[}62.28, 82.73{]} & 65.65   {[}56.85, 73.72{]} & 56.31   {[}49.56, 62.84{]} & 80.37   {[}73.56, 85.77{]} \\
 & I-T & 0.739   {[}0.675, 0.796{]} & 69.27   {[}62.68, 75.32{]} & 85.71   {[}76.38, 92.39{]} & 58.96   {[}50.13, 67.37{]} & 56.69   {[}51.21, 62.02{]} & 86.81   {[}79.28, 91.89{]} \\
\midrule
\multirow{3}{*}{Inception-v3} & T & 0.968 {[}0.951, 0.980{]} & 91.75 {[}89.32, 93.77{]} & 95.42 {[}91.95, 97.69{]} & 89.49 {[}86.01, 92.35{]} & 84.81 {[}80.68, 88.20{]} & 96.94 {[}94.68, 98.26{]} \\
 & V & 0.753 {[}0.689, 0.810{]} & 70.48 {[}63.81, 76.55{]} & 67.09 {[}55.60, 77.25{]} & 72.52 {[}64.04, 79.95{]} & 59.55 {[}51.71, 66.93{]} & 78.51 {[}72.39, 83.59{]} \\
 & I-T & 0.762 {[}0.700, 0.817{]} & 71.10 {[}64.59, 77.02{]} & 85.71 {[}76.38, 92.39{]} & 61.94 {[}53.16, 70.18{]} & 58.54 {[}52.79, 64.06{]} & 87.37 {[}80.12, 92.23{]} \\
\midrule
\multirow{3}{*}{VGG16\_BN} & T & 0.901   {[}0.875, 0.923{]} & 80.32   {[}76.99, 83.35{]} & 94.17   {[}90.41, 96.77{]} & 71.79   {[}67.05, 76.21{]} & 67.26   {[}63.61, 70.71{]} & 95.24   {[}92.30, 97.09{]} \\
 & V & 0.808   {[}0.748, 0.859{]} & 72.86   {[}66.31, 78.75{]} & 77.22   {[}66.40, 85.90{]} & 70.23   {[}61.62, 77.90{]} & 61.00   {[}53.95, 67.62{]} & 83.64   {[}77.04, 88.62{]} \\
 & I-T & 0.816   {[}0.758, 0.865{]} & 74.77   {[}68.46, 80.39{]} & 80.95   {[}70.92, 88.70{]} & 70.90   {[}62.43, 78.42{]} & 63.55   {[}56.76, 69.84{]} & 85.59   {[}79.04, 90.34{]} \\
\midrule
\multirow{3}{*}{VGG16\_BN+C} & T & 0.878   {[}0.622, 0.698{]} & 76.51   {[}73.00, 79.77{]} & 93.33   {[}89.40, 96.14{]} & 66.15   {[}61.22, 70.84{]} & 62.92   {[}59.53, 66.19{]} & 94.16   {[}90.90, 96.30{]} \\
 & V & 0.823   {[}0.765, 0.872{]} & 75.71   {[}69.34, 81.35{]} & 74.68   {[}63.64, 83.80{]} & 76.34   {[}68.12, 83.32{]} & 65.56   {[}57.69, 72.65{]} & 83.33   {[}77.19, 88.08{]} \\
 & I-T & 0.831   {[}0.775, 0.878{]} & 75.69   {[}69.44, 81.23{]} & 89.29   {[}80.63, 94.98{]} & 67.16   {[}58.53, 75.03{]} & 63.03   {[}56.96, 68.71{]} & 90.91   {[}84.21, 94.94{]} \\
\bottomrule
\end{tabular}
\begin{tablenotes}
\footnotesize
\item 95\% confidence intervals are included in brackets.
\item AUC area under the receiver operating characteristic curve, ACC accuracy, SENS sensitivity, SPEC specificity, PPV positive predict value, NPV negative predict value.
\item T training cohort ($n$ = 630), V validation cohort ($n$ = 210), I–T independent test cohort ($n$ = 218).
\end{tablenotes}
\end{threeparttable}
}
\end{table}

\begin{table}[htb]
\centering
\caption{The detailed parameters of VGG16\_BN.}
\label{tab:table_s_2}
\resizebox{0.8\linewidth}{!}{
\renewcommand\arraystretch{1.3}
\begin{threeparttable}
\begin{tabular}{lllllll}
\toprule
Layer name & Input channels & Output channels & Kernel size & Stride & Padding & Output size \\
\midrule
basic block $\times$ 2 & 3 & 64 & 3 & 1 & 1 & {[}64, 256, 256{]} \\
max-pooling layer &  &  & 2 & 2 & 0 & {[}64, 128, 128{]} \\
basic block $\times$ 2 & 64 & 128 & 3 & 1 & 1 & {[}128, 128, 128{]} \\
max-pooling layer &  &  & 2 & 2 & 0 & {[}128, 64, 64{]} \\
basic block $\times$ 3 & 128 & 256 & 3 & 1 & 1 & {[}256, 64, 64{]} \\
max-pooling layer &  &  & 2 & 2 & 0 & {[}256, 32, 32{]} \\
basic block $\times$ 3 & 256 & 512 & 3 & 1 & 1 & {[}512, 32, 32{]} \\
max-pooling layer &  &  & 2 & 2 & 0 & {[}512, 16, 16{]} \\
basic block $\times$ 3 & 512 & 512 & 3 & 1 & 1 & {[}512, 16, 16{]} \\
max-pooling layer &  &  & 2 & 2 & 0 & {[}512, 8, 8{]} \\
adaptive average pooling   layer &  &  &  &  &  & {[}512, 7, 7{]} \\
\bottomrule
\end{tabular}
\begin{tablenotes}
\footnotesize
\item The basic block was cascade by convolution layer, batch normalization layer, and Rectified Linear Unit (ReLU).
\item The input size of the model was [3, 256, 256], which followed the format of [channel, height, width].
\end{tablenotes}
\end{threeparttable}
}
\end{table}

\begin{table}[htb]
\centering
\caption{The performance in prediction of ALN status (N0 vs. N$_+$(1-2)).}
\label{tab:table_s_3}
\resizebox{1.0\linewidth}{!}{
\renewcommand\arraystretch{1.3}
\begin{threeparttable}
\begin{tabular}{llllllll}
\toprule
Methods &  & AUC & ACC (\%) & SENS (\%) & SPEC (\%) & PPV (\%) & NPV (\%) \\
\midrule
\multirow{3}{*}{Clinical data   only} & T & 0.638 {[}0.595, 0.679{]} & 61.00 {[}56.65, 65.23{]} & 65.62 {[}56.72, 73.79{]} & 59.49 {[}54.43, 64.40{]} & 34.71 {[}30.88, 38.75{]} & 84.06 {[}80.37, 87.16{]} \\
 & V & 0.677 {[}0.602, 0.745{]} & 74.29 {[}67.15, 80.58{]} & 45.45 {[}30.39, 61.15{]} & 83.97 {[}76.55, 89.79{]} & 48.78 {[}36.42, 61.29{]} & 82.09 {[}77.60, 85.84{]} \\
 & I-T & 0.627$^{a, b}$ {[}0.551,   0.700{]} & 72.67 {[}65.37, 79.18{]} & 44.74 {[}28.62, 61.70{]} & 80.60 {[}72.88, 86.92{]} & 39.53 {[}28.52, 51.73{]} & 83.72 {[}79.24, 87.39{]} \\
\midrule
\multirow{3}{*}{DL-CNB model} & T & 0.912 {[}0.884, 0.935{]} & 82.24 {[}78.67, 85.44{]} & 97.66 {[}93.30, 99.51{]} & 77.18 {[}72.69, 81.25{]} & 58.41 {[}53.87, 62.81{]} & 99.01 {[}97.04, 99.68{]} \\
 & V & 0.756 {[}0.685, 0.817{]} & 59.43 {[}51.76, 66.77{]} & 97.73 {[}87.98, 99.94{]} & 46.56 {[}37.81, 55.48{]} & 38.05 {[}34.22, 42.04{]} & 98.39 {[}89.70, 99.77{]} \\
 & I-T & 0.845$^{c}$ {[}0.782,   0.895{]} & 80.23 {[}73.49, 85.90{]} & 73.68 {[}56.90, 86.60{]} & 82.09 {[}74.53, 88.17{]} & 53.85 {[}43.66, 63.72{]} & 91.67 {[}86.53, 94.96{]} \\
\midrule
\multirow{3}{*}{DL-CNB+C model} & T & 0.936 {[}0.911, 0.955{]} & 84.17 {[}80.74, 87.21{]} & 95.31 {[}90.08, 98.26{]} & 80.51 {[}76.23, 84.33{]} & 61.62 {[}56.66, 66.34{]} & 98.12 {[}95.99, 99.13{]} \\
 & V & 0.789 {[}0.721, 0.847{]} & 66.29 {[}58.76, 73.24{]} & 84.09 {[}69.93, 93.36{]} & 60.31 {[}51.39, 68.74{]} & 41.57 {[}35.72, 47.67{]} & 91.86 {[}84.94, 95.76{]} \\
 & I-T & 0.878 {[}0.819, 0.923{]} & 84.30 {[}77.99, 89.39{]} & 71.05 {[}54.10, 84.58{]} & 88.06 {[}81.33, 93.02{]} & 62.79 {[}50.52, 73.61{]} & 91.47 {[}86.65, 94.66{]} \\
\bottomrule
\end{tabular}
\begin{tablenotes}
\footnotesize
\item 95\% confidence intervals are included in brackets.
\item AUC, area under the receiver operating characteristic curve; ACC, accuracy; SENS, sensitivity; SPEC, specificity; PPV, positive predictive value; NPV, negative predictive value.
\item T, training cohort ($n$ = 518); V, validation cohort ($n$ = 175); I–T, independent test cohort ($n$ = 172).
\item ALN, axillary lymph node; DL-CNB+C, deep learning core-needle biopsy incorporating the clinical data.
\item $^{a}$ Indicates $p$ = 0.0004, Delong et al. in comparison with DL-CNB model in independent test cohort.
\item $^{b}$ Indicates $p$ < 0.0001, Delong et al. in comparison with DL-CNB+C model in independent test cohort.
\item $^{c}$ Indicates $p$ = 0.1148, Delong et al. in comparison with DL-CNB+C model in independent test cohort.
\end{tablenotes}
\end{threeparttable}
}
\end{table}

\begin{table}[htb]
\centering
\caption{The performance in prediction of ALN status (N0 vs. N$_+$($\geq$3)).}
\label{tab:table_s_4}
\resizebox{1.0\linewidth}{!}{
\renewcommand\arraystretch{1.3}
\begin{threeparttable}
\begin{tabular}{llllllll}
\toprule
Methods &  & AUC & ACC (\%) & SENS (\%) & SPEC (\%) & PPV (\%) & NPV (\%) \\
\midrule
\multirow{3}{*}{Clinical data   only} & T & 0.680 {[}0.638, 0.721{]} & 66.67 {[}62.39, 70.75{]} & 65.83 {[}56.62, 74.24{]} & 66.92 {[}62.01, 71.58{]} & 37.98 {[}33.59, 42.58{]} & 86.42 {[}83.10, 89.18{]} \\
 & V & 0.748 {[}0.675, 0.813{]} & 71.52 {[}63.98, 78.26{]} & 76.47 {[}58.83, 89.25{]} & 70.23 {[}61.62, 77.90{]} & 40.00 {[}32.57, 47.92{]} & 92.00 {[}86.13, 95.51{]} \\
 & I-T & 0.629$^{a, b}$ {[}0.553,   0.701{]} & 69.36 {[}61.92, 76.14{]} & 53.85 {[}37.18, 69.91{]} & 73.88 {[}65.59, 81.08{]} & 37.50 {[}28.54, 47.40{]} & 84.62 {[}79.43, 88.68{]} \\
\midrule
\multirow{3}{*}{DL-CNB model} & T & 0.906 {[}0.877, 0.930{]} & 81.57 {[}77.93, 84.84{]} & 93.33 {[}87.29, 97.08{]} & 77.95 {[}73.50, 81.97{]} & 56.57 {[}51.79, 61.23{]} & 97.44 {[}95.10, 98.67{]} \\
 & V & 0.755 {[}0.682, 0.819{]} & 64.24 {[}56.42, 71.54{]} & 91.18 {[}76.32, 98.14{]} & 57.25 {[}48.32, 65.85{]} & 35.63 {[}30.67, 40.92{]} & 96.15 {[}89.36, 98.67{]} \\
 & I-T & 0.837$^{c}$ {[}0.773,   0.888{]} & 69.94 {[}62.52, 76.67{]} & 92.31 {[}79.13, 98.38{]} & 63.43 {[}54.68, 71.58{]} & 42.35 {[}36.61, 48.31{]} & 96.59 {[}90.46, 98.83{]} \\
\midrule
\multirow{3}{*}{DL-CNB+C model} & T & 0.918 {[}0.891, 0.940{]} & 82.16 {[}78.55, 85.38{]} & 91.67 {[}85.21, 95.93{]} & 79.23 {[}74.86, 83.15{]} & 57.59 {[}52.62, 62.42{]} & 96.87 {[}94.45, 98.25{]} \\
 & V & 0.761 {[}0.689, 0.824{]} & 66.06 {[}58.29, 73.24{]} & 79.41 {[}62.10, 91.30{]} & 62.60 {[}53.72, 70.89{]} & 35.53 {[}29.40, 42.16{]} & 92.13 {[}85.66, 95.83{]} \\
 & I-T & 0.838 {[}0.774, 0.889{]} & 71.10 {[}63.73, 77.73{]} & 89.74 {[}75.78, 97.13{]} & 65.67 {[}56.98, 73.65{]} & 43.21 {[}37.04, 49.60{]} & 95.65 {[}89.61, 98.25{]} \\
\bottomrule
\end{tabular}
\begin{tablenotes}
\footnotesize
\item 95\% confidence intervals are included in brackets.
\item AUC, area under the receiver operating characteristic curve; ACC, accuracy; SENS, sensitivity; SPEC, specificity; PPV, positive predictive value; NPV, negative predictive value.
\item T, training cohort ($n$ = 510); V, validation cohort ($n$ = 165); I–T, independent test cohort ($n$ = 173).
\item ALN, axillary lymph node; DL-CNB+C, deep learning core-needle biopsy incorporating the clinical data.
\item $^{a}$ Indicates $p$ = 0.0005, Delong et al. in comparison with DL-CNB model in independent test cohort.
\item $^{b}$ Indicates $p$ < 0.0001, Delong et al. in comparison with DL-CNB+C model in independent test cohort.
\item $^{c}$ Indicates $p$ = 0.9689, Delong et al. in comparison with DL-CNB+C model in independent test cohort.
\end{tablenotes}
\end{threeparttable}
}
\end{table}

\begin{table}[htb]
\centering
\caption{The subgroup performance in prediction of ALN status by DL-CNB+C model (N0 vs. N(+)).}
\label{tab:table_s_5}
\resizebox{1.0\linewidth}{!}{
\renewcommand\arraystretch{1.3}
\begin{threeparttable}
\begin{tabular}{llllllllll}
\toprule
Characteristics & Value &  & AUC & ACC (\%) & SENS (\%) & SPEC (\%) & PPV (\%) & NPV (\%) & $p$ \\
\midrule
\multirow{2}{*}{Age $\leq$ 50} & Yes & I-T & 0.918 {[}0.825, 0.971{]} & 82.09 {[}70.80, 90.39{]} & 93.33 {[}77.93, 99.18{]} & 72.97 {[}55.88, 86.21{]} & 73.68 {[}62.05, 82.74{]} & 93.10 {[}77.72, 98.12{]} & \multirow{2}{*}{0.0151} \\
 & No & I-T & 0.794 {[}0.720, 0.855{]} & 66.89 {[}58.77, 74.32{]} & 90.74 {[}79.70, 96.92{]} & 53.61 {[}43.19, 63.8{]} & 52.13 {[}46.38, 57.82{]} & 91.23 {[}81.56, 96.07{]} &  \\
\midrule
\multirow{2}{*}{T stage} & T1 & I-T & 0.833 {[}0.754, 0.895{]} & 71.90 {[}63.01, 79.69{]} & 89.19 {[}74.58, 96.97{]} & 64.29 {[}53.08, 74.45{]} & 52.38 {[}44.70, 59.95{]} & 93.10 {[}84.07, 97.19{]} & \multirow{2}{*}{0.7426} \\
 & T2 & I-T & 0.814 {[}0.722, 0.886{]} & 71.13 {[}61.05, 79.89{]} & 93.62 {[}82.46, 98.66{]} & 50.00 {[}35.53, 64.47{]} & 63.77 {[}56.91, 70.11{]} & 89.29 {[}72.93, 96.27{]} &  \\
\midrule
\multirow{2}{*}{ER} & Positive & I-T & 0.853 {[}0.789, 0.903{]} & 82.53 {[}75.88, 87.98{]} & 89.23 {[}79.06, 95.56{]} & 78.22 {[}68.90, 85.82{]} & 72.50 {[}64.34, 79.39{]} & 91.86 {[}84.76, 95.81{]} & \multirow{2}{*}{0.1253} \\
 & Negative & I-T & 0.737 {[}0.596, 0.849{]} & 67.31 {[}52.89, 79.67{]} & 73.68 {[}48.80, 90.85{]} & 63.64 {[}45.12, 79.60{]} & 53.85 {[}40.83, 66.36{]} & 80.77 {[}65.47, 90.30{]} &  \\
\midrule
\multirow{2}{*}{PR} & Positive & I-T & 0.839 {[}0.772, 0.893{]} & 72.61 {[}64.93, 79.42{]} & 96.61 {[}88.29, 99.59{]} & 58.16 {[}47.77, 68.05{]} & 58.16 {[}52.28, 63.82{]} & 96.61 {[}87.84, 99.12{]} & \multirow{2}{*}{0.6591} \\
 & Negative & I-T & 0.811 {[}0.690, 0.900{]} & 70.49 {[}57.43, 81.48{]} & 80.00 {[}59.30, 93.17{]} & 63.89 {[}46.22, 79.18{]} & 60.61 {[}48.85, 71.25{]} & 82.14 {[}66.92, 91.27{]} &  \\
\midrule
\multirow{2}{*}{HER2} & Positive & I-T & 0.800 {[}0.677, 0.892{]} & 66.67 {[}53.31, 78.31{]} & 88.89 {[}65.29, 98.62{]} & 57.14 {[}40.96, 72.28{]} & 47.06 {[}37.68, 56.65{]} & 92.31 {[}75.99, 97.85{]} & \multirow{2}{*}{0.5238} \\
 & Negative & I-T & 0.842 {[}0.776, 0.895{]} & 74.05 {[}66.49, 80.69{]} & 92.42 {[}83.20, 97.49{]} & 60.87 {[}50.14, 70.88{]} & 62.89 {[}56.54, 68.81{]} & 91.80 {[}82.60, 96.35{]} &  \\
\bottomrule
\end{tabular}
\begin{tablenotes}
\footnotesize
\item 95\% confidence intervals are included in brackets.
\item AUC, area under the receiver operating characteristic curve; ACC, accuracy; SENS, sensitivity; SPEC, specificity; PPV, positive predictive value; NPV, negative predictive value.
\item I-T independent test group, ER estrogen receptor, PR progesterone receptor, HER-2 human epidermal growth factor receptor-2, LNM lymph node metastasis.
\end{tablenotes}
\end{threeparttable}
}
\end{table}

\begin{figure}[htb]
\centering
\includegraphics[width=0.6\linewidth]{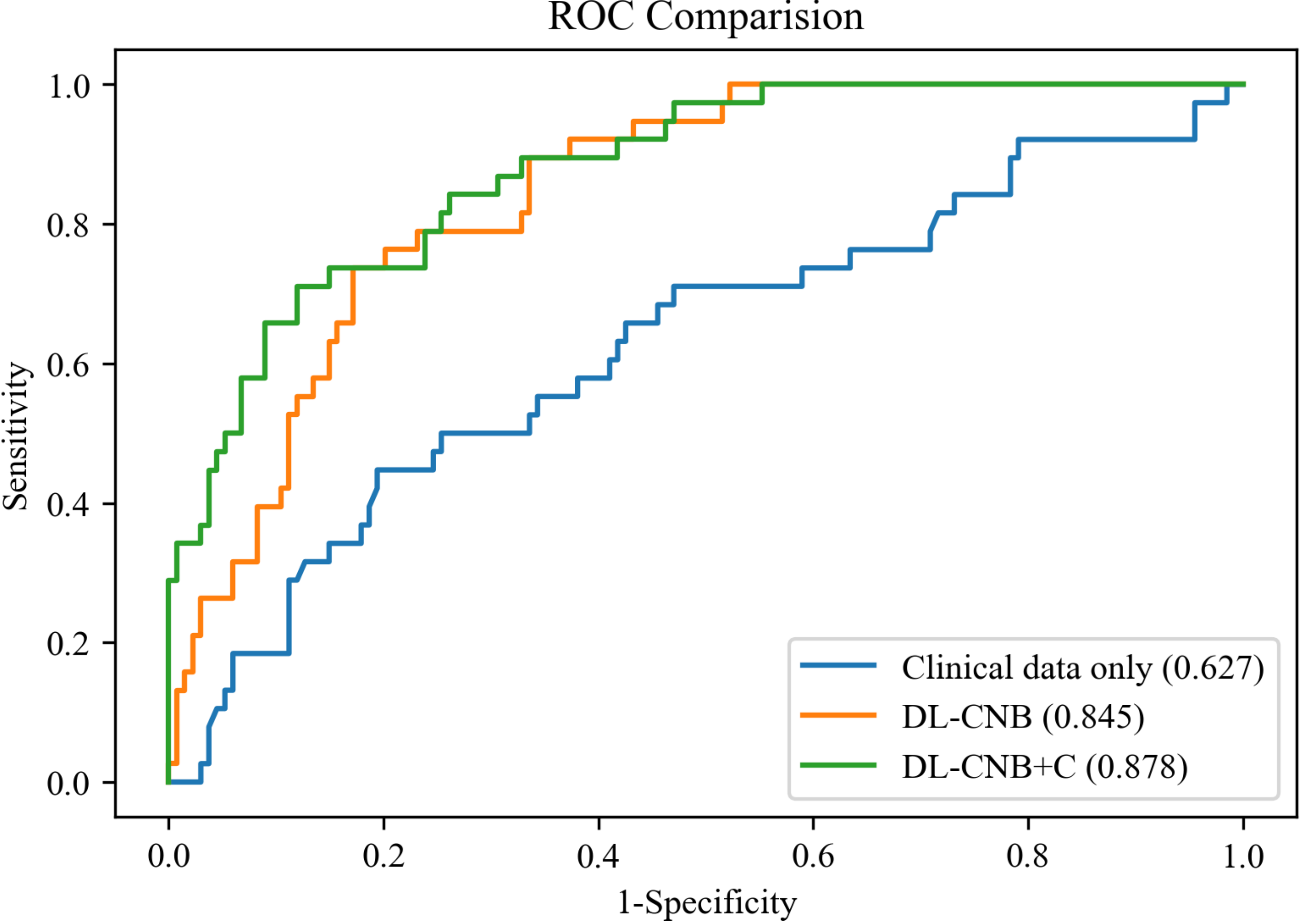}
\caption{Comparison of receiver operating characteristic (ROC) curves between different models for predicting disease-free axilla (N0) and low metastatic burden of axillary disease (N$_+$(1-2)). Numbers in parentheses are areas under the receiver operating characteristic curve (AUCs).}
\label{fig:fig_s_1}
\end{figure}

\begin{figure}[t!]
\centering
\includegraphics[width=0.6\linewidth]{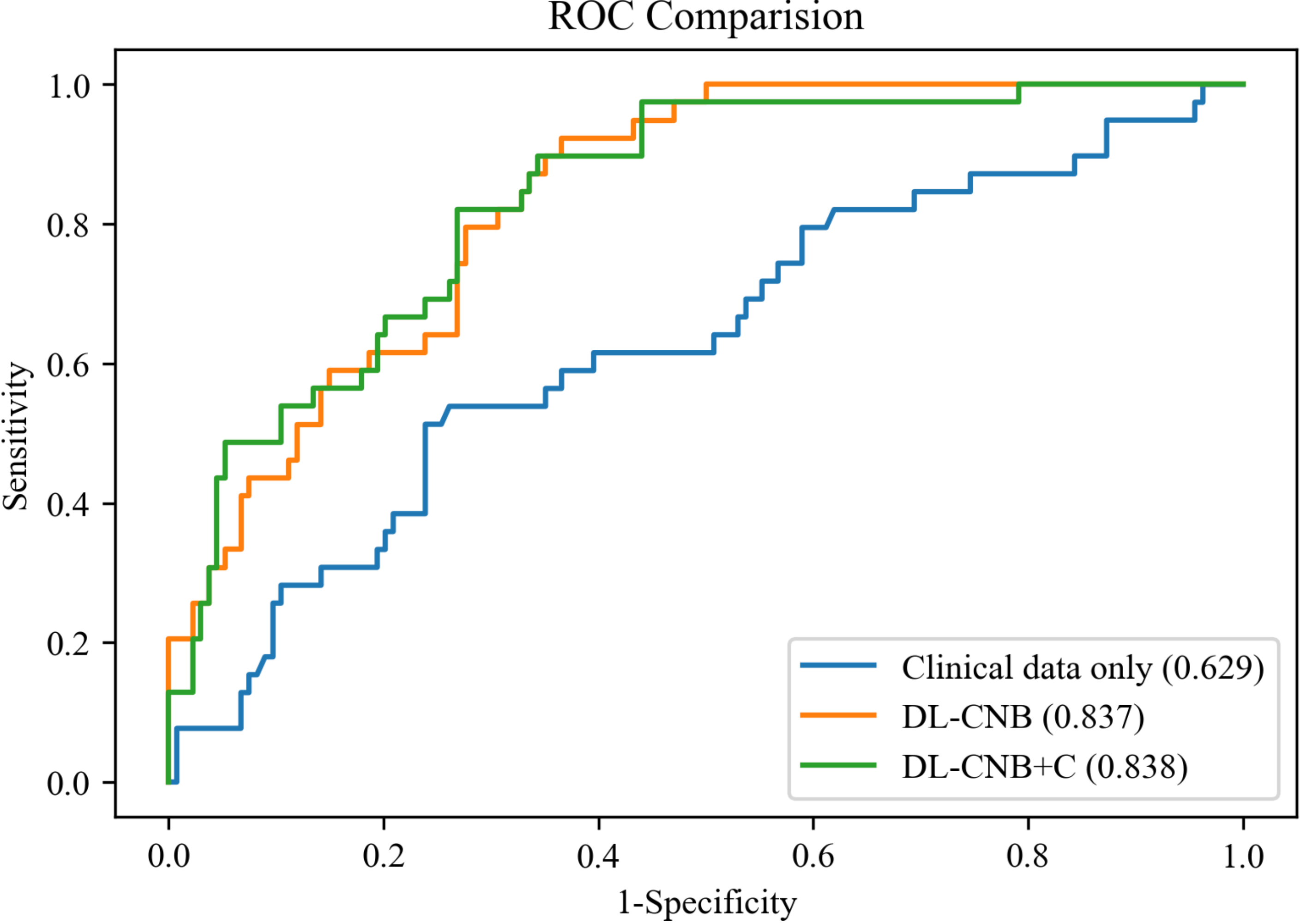}
\caption{Comparison of receiver operating characteristic (ROC) curves between different models for predicting disease-free axilla (N0) and low metastatic burden of axillary disease (N$_+$($\geq$3)). Numbers in parentheses are areas under the receiver operating characteristic curve (AUCs).}
\label{fig:fig_s_2}
\end{figure}